\documentstyle [12pt] {article}
\begin{document}

\title{
On Possibility of Crystal Extraction and Collimation\\ at 0.1-1 GeV}
\author{
V.~M.~Biryukov\\
\em Institute for High Energy Physics, Protvino, Russia}

\date{PAC 1999 Proceedings (New York), pp.1240-1242}

\maketitle

\begin{abstract}
Bent crystal situated in a circulating beam can serve for efficient
slow extraction or active collimation of the beams.
This technique, well established at 10-1000 GeV, could be efficient
also at the energies as low as 0.1-10 GeV according to the computer
simulations presented in this paper. Applications might include
halo scraping in the Spallation Neutron Source or slow extraction
from synchrotrons.
\end{abstract}

\section{Introduction}

Bent crystals are successfully applied at IHEP Protvino\cite{ihep}
for slow extraction of 70 GeV protons.
This technique is well explored in a broad energy range 14-900 GeV
\cite{book},
and some projects for halo scraping with crystals at RHIC and Tevatron
are in progress\cite{rhic,fnal}.
It would be interesting to study the feasibility of this technique also
at essentially lower energies, where it may assist in scraping beam halos
e.g. at the Spallation Neutron Source, or assist in  slow extraction from
multi-hundred MeV proton or light ion accelerators\cite{walter}.
In the present contribution we study only physical aspects of the job,
i.e. how much of the beam can be steered at how much angle.

Bent crystal channeling was first demonstrated at a few GeV in Dubna, 1979,
and its physics first studied at 1 to 12 GeV at Petersburg, Dubna and CERN.
Crystal-based extraction from accelerator was also first demonstrated
at 4-8 GeV in Dubna in 1984.
So the 1-10 GeV domain is well familiar to bent crystals.

\section{Crystal Collimation at the SNS}

The Spallation Neutron Source will require an efficient halo collimation
of the protons accumulated in 1 GeV ring.
The general idea of the crystal-assisted collimation
is that a  crystal, serving as a primary element,
gives the incident halo particles a bend of e.g.
a few mrad so as to provide a big impact parameter at some secondary element.
Then the bent particles are absorbed with a higher probability there, and so
the backscattering is much less a problem. A radical solution might even be
an extraction of bent particles to some external dump.

\underline{\em "Single-pass scraping"}.
A straightforward option would be to bend particles in a single pass
through the bent crystal, at an angle of the order of 10 mrad.
The protons with 1 GeV kinetic energy can move through Silicon
in the channeled states over 1-1.5 mm. The bending radius $R$
must be greater than critical one ($R_c$=0.25 cm here);
we take $R$=10 cm as an example for our simulation.
Figure 1 shows the angular distribution downstream of the bent
crystal 1 mm long.
In this simulation first we assumed that proton divergence at the incidence
on the crystal was much narrower than
the acceptance of the
Si(110) crystal planes 2$\theta_c$=0.25 mrad.
About 77\% of the particles have exit angles greater than 1 mrad,
and 30\% of the total are bent at the full bending angle, 10 mrad.
Further example in Figure 1 repeats the simulation for the incident
beam with divergence of 0.2 mrad.
In that case 70\% of the particles are bent more than 1 mrad,
and 22\% of the total are bent at 10 mrad.

\begin{figure}[htb]
\begin{center}
\setlength{\unitlength}{.45mm}
\begin{picture}(120,100)(-10,-3)
\thicklines
\linethickness{.5mm}
\put(    -20. ,1.8)  {\line(1,0){5}}
\put(    -15. ,2.4)  {\line(1,0){5}}
\put(    -10. ,6.0)  {\line(0,-1){3.6}}
\put(    -10. ,6.0)  {\line(1,0){5}}
\put(     -5. ,6.8)  {\line(1,0){5}}
\put(      0.,11.4)  {\line(0,-1){4.6}}
\put(      0.,11.4)  {\line(1,0){5}}
\put(      5.,11.6)  {\line(1,0){5}}
\put(     10. ,8.4)  {\line(0,1){3.2}}
\put(     10. ,8.4)  {\line(1,0){5}}
\put(     15. ,8.4)  {\line(1,0){5}}
\put(     20. ,8.8)  {\line(1,0){5}}
\put(     25. ,7.8)  {\line(1,0){5}}
\put(     30. ,6.6)  {\line(1,0){5}}
\put(     35. ,3.8)  {\line(1,0){5}}
\put(     40. ,6.2)  {\line(1,0){5}}
\put(     45. ,5.2)  {\line(1,0){5}}
\put(     50. ,6.6)  {\line(1,0){5}}
\put(     55. ,4.6)  {\line(1,0){5}}
\put(     60. ,6.0)  {\line(1,0){5}}
\put(     65. ,3.4)  {\line(1,0){5}}
\put(     70. ,4.6)  {\line(1,0){5}}
\put(     75. ,3.4)  {\line(1,0){5}}
\put(     80. ,5.6)  {\line(1,0){5}}
\put(     85. ,4.6)  {\line(1,0){5}}
\put(     90. ,3.4)  {\line(1,0){5}}
\put(     95. ,3.6)  {\line(1,0){5}}
\put(    100. ,59.0)  {\line(0,-1){55.4}}
\put(    100. ,59.0)  {\line(1,0){5}}
\put(    105. ,59.0)  {\line(0,-1){59}}

\linethickness{.15mm}
\put(    -20. ,1.2)  {\line(1,0){5}}
\put(    -15. ,4.6)  {\line(0,-1){3.4}}
\put(    -15. ,4.6)  {\line(1,0){5}}
\put(    -10. ,7.4)  {\line(1,0){5}}
\put(     -5. ,8.)  {\line(1,0){5}}
\put(      0. ,15.0)  {\line(0,-1){7}}
\put(      0. ,15.0)  {\line(1,0){5}}
\put(      5. ,16.2)  {\line(0,-1){1.2}}
\put(      5. ,16.2)  {\line(1,0){5}}
\put(     10. ,16.2)  {\line(0,-1){1.6}}
\put(     10. ,14.6)  {\line(1,0){5}}
\put(     15. ,14.6)  {\line(0,-1){5.2}}
\put(     15. ,9.4)  {\line(1,0){5}}
\put(     20. ,8.2)  {\line(1,0){5}}
\put(     25. ,8.2)  {\line(1,0){5}}
\put(     30. ,7.2)  {\line(1,0){5}}
\put(     35. ,7.6)  {\line(1,0){5}}
\put(     40. ,7.4)  {\line(1,0){5}}
\put(     45. ,5.2)  {\line(1,0){5}}
\put(     50. ,6.4)  {\line(1,0){5}}
\put(     55. ,5.0)  {\line(1,0){5}}
\put(     60. ,2.8)  {\line(1,0){5}}
\put(     65. ,3.6)  {\line(1,0){5}}
\put(     70. ,3.8)  {\line(1,0){5}}
\put(     75. ,3.2)  {\line(1,0){5}}
\put(     80. ,4.0)  {\line(1,0){5}}
\put(     85. ,2.2)  {\line(1,0){5}}
\put(     90. ,3.0)  {\line(1,0){5}}
\put(     95. ,1.6)  {\line(1,0){5}}
\put(    100. ,43.6)  {\line(0,-1){43.6}}
\put(    100. ,43.6)  {\line(1,0){5}}
\put(    105. ,43.6)  {\line(0,-1){43.6}}
\linethickness{.25mm}
\put(-20,0) {\line(1,0){140}}
\put(-20,0) {\line(0,1){100}}
\put(-20,100) {\line(1,0){140}}
\put(120,0){\line(0,1){100}}
\multiput( 2.5,0)(50,0){3}{\line(0,1){3}}
\multiput(-7.5,0)(10,0){12}{\line(0,1){1.4}}
\multiput(-20,20)(0,20){4}{\line(1,0){3}}
\multiput(-20,4)(0,4){25}{\line(1,0){1.}}
\put(2.5,-8){\makebox(1,1)[b]{0}}
\put(52.5,-8){\makebox(1,1)[b]{5}}
\put(102.5,-8){\makebox(1,1)[b]{10}}
\put(-17,20){\makebox(1,.5)[l]{10}}
\put(-17,40){\makebox(1,.5)[l]{20}}
\put(-17,60){\makebox(1,.5)[l]{30}}
\put(-17,80){\makebox(1,.5)[l]{40}}

\put(-17,90){\large N(rel.)}
\put(60,-15){\large Angle (mrad)}

\end{picture}
\end{center}
\caption{
Angular distribution of 1 GeV protons downstream of the
silicon crystal bent 10 mrad.
For the parallel incident beam (thick line),
anf for the beam incident with divergence of 0.2 mrad (thin line).
}
  \label{single}
\end{figure}
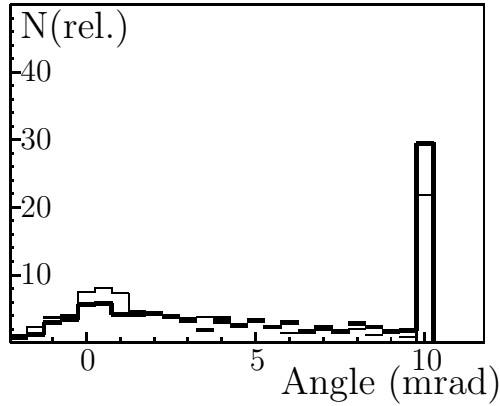

Actual bending angles and the particle distribution downstream
of a bent crystal can be designed as required to match some
particular design of a cleaning system. The examples shown are for
illustrative purpose only.

\underline{\em "Multi-pass scraping"}.
Rather interesting would the option where halo particles can encounter
the crystal several times (multiturn, multipass 'extraction'),
while circulating in the ring,
which increases the extraction efficiency substantially.
This option is feasible if the crystal is short enough along the beam,
to reduce particle losses and scattering when it encounters the crystal,
thus retaining the scattered particles in circulation.

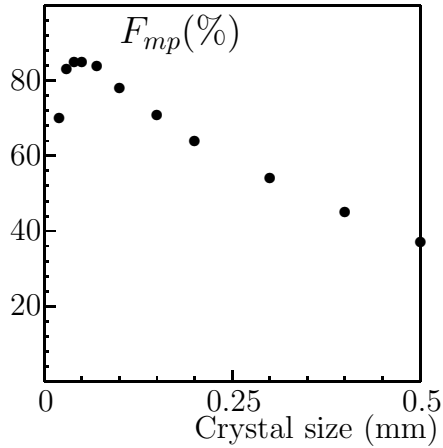
\begin{figure}[htb]
\begin{center}
\setlength{\unitlength}{.5mm}
\begin{picture}(100,100)(0,-3)
\thicklines

\put( 4. ,70) {\circle*{3}}
\put( 6. ,83) {\circle*{3}}
\put( 8. ,85) {\circle*{3}}
\put( 10. ,85) {\circle*{3}}
\put( 14. ,84) {\circle*{3}}
\put( 20. ,78) {\circle*{3}}
\put( 30. ,71) {\circle*{3}}
\put( 40. ,64) {\circle*{3}}
\put( 60. ,54) {\circle*{3}}
\put( 80. ,45) {\circle*{3}}
\put( 100. ,37) {\circle*{3}}

\linethickness{.25mm}
\put(0,0) {\line(1,0){100}}
\put(0,0) {\line(0,1){100}}
\put(0,100) {\line(1,0){100}}
\put(100,0){\line(0,1){100}}
\multiput(0,0)(50,0){3}{\line(0,1){3}}
\multiput(0,0)(10,0){11}{\line(0,1){1.4}}
\multiput(0,20)(0,20){4}{\line(1,0){3}}
\multiput(0,4)(0,4){25}{\line(1,0){1.}}
\put(0,-8){\makebox(1,1)[b]{0}}
\put(50,-8){\makebox(1,1)[b]{0.25}}
\put(100,-8){\makebox(1,1)[b]{0.5}}
\put(-9,20){\makebox(1,.5)[l]{20}}
\put(-9,40){\makebox(1,.5)[l]{40}}
\put(-9,60){\makebox(1,.5)[l]{60}}
\put(-9,80){\makebox(1,.5)[l]{80}}

\put(20,90){\large $F_{mp}$(\%)}
\put(40,-15){Crystal size (mm)}

\end{picture}
\end{center}
\caption{
Efficiency of crystal multipass scraping
for the 1 GeV particles that were not channeled on the first encounter,
as a function of the crystal size along the beam.
}
  \label{multi}
\end{figure}

In this option, several bending angles were tried from a fraction of mrad up
to 5 mrad. Upon first encounter with a crystal, particles were allowed to
circulate in the SNS ring (linear transfer matrice was used
with $\nu_{x,y}$=5.82/5.80 and beta tentatively chosen as 10 m)
and have further encounters with crystal or
aperture (in that case the particle was removed). The aperture limitation
was imposed on the particle's angle at the crystal location: if its absolute
value was greater than 0.5 (but less than 0.9) of the crystal bending angle,
that particle was removed (considered lost at a collimator edge),
so we counted only channeled particles that were steered at the proper angle.

The efficiency (the number of particles bent the full angle) was
roughly independent
of the chosen bending angle in the range studied, up to 5 mrad,
and totalled typically 85-90\% of the beam incident on the crystal.
The unchanneled 10-15\% of the particles reach the collimator edge
(the aperture) and can be handled by more traditional "amorphous"
collimation.

Figure 2 shows the probability with which the particles unchanneled
in the first encounter with crystal (bent 3 mrad)
are channeled on later encounters.
If the crystal is as short along the beam as order of 50 $\mu$m,
this probability is very high, so even the particles multiply scattered
in the crystal at first incidence can be efficiently channeled
on later turns and steered away.
Due to multiple encounters,
the initial divergence of particles at the crystal becomes not so critical
and can be about the scattering angle along the crystal length.
Crystal efficiency in multi-pass mode
is defined mainly by the interplay of channeling and scattering processes in
multiple encounters with a short crystal.
The overall energy loss in multiple encounters with crystal is within
the nominal energy spread of the SNS beam,
$\Delta E/E<$4$\times$10$^{-3}$ (rms).

For a typical particle, it takes 5 to 10 encounters
with a crystal on average
before the particle is channeled and extracted.
This corresponds to order of 100
turns from the moment of the first encounter with
a crystal to the moment of the particle extraction
from circulation in the ring.

Further simulation involving realistic description of the SNS
machine and beam parameters will be necessary for realistic
evaluation of a crystal-assisted scraping.

\section{Slow extraction from multi-hundred MeV machines}

The above-considered multipass channeling
may be well suited for efficient slow extraction from
medium-energy synchrotrons \cite{walter}.
The analytical theory\cite{theory} expects that
as the beam energy lowers from multi-GeV to 0.1-1
GeV, one could reduce by a big factor the crystal size. Tiny
crystal size may permit a huge multiplicity of particle encounters
with the crystal, and hence a very high overall efficiency of crystal
channeling.

\begin{figure}[htb]
\begin{center}
\setlength{\unitlength}{.5mm}
\begin{picture}(100,100)(0,-3)
\thicklines

\put( 10 ,73) {\circle*{3}}      
\put( 22. ,77) {\circle*{3}}
\put( 30. ,79) {\circle*{3}}     
\put( 38. ,81) {\circle*{3}}
\put( 50. ,84) {\circle*{3}}
\put( 70. ,85) {\circle*{3}}
\put( 90. ,83) {\circle*{3}}

\put( 10 ,-1) {\large $\star$}      
\put( 22 ,-1) {\large $\star$}
\put( 30 ,-0.8) {\large $\star$}
\put( 38 ,-0.7) {\large $\star$}
\put( 50 ,-0.4) {\large $\star$}
\put( 60 ,-0.15) {\large $\star$}
\put( 70 ,0.2) {\large $\star$}
\put( 77 ,0.9) {\large $\star$}
\put( 83 ,3.2) {\large $\star$}
\put( 90 ,6.3) {\large $\star$}      

\put( 10 ,20) {$\otimes$}
\put( 22 ,15) {$\otimes$}
\put( 30 ,12) {$\otimes$}
\put( 38 ,11) {$\otimes$}
\put( 50 ,7) {$\otimes$}
\put( 70 ,5) {$\otimes$}
\put( 77 ,4) {$\otimes$}
\put( 90 ,1.8) {$\otimes$}

\linethickness{.25mm}
\put(0,0) {\line(1,0){100}}
\put(0,0) {\line(0,1){100}}
\put(0,100) {\line(1,0){100}}
\put(100,0){\line(0,1){100}}
\multiput(10,0)(40,0){3}{\line(0,1){3}}
\multiput(10,0)(20,0){5}{\line(0,1){1.4}}
\multiput(0,20)(0,20){4}{\line(1,0){3}}
\multiput(0,4)(0,4){25}{\line(1,0){1.}}
\put(10,-8){\makebox(1,1)[b]{0.1}}
\put(30,-8){\makebox(1,1)[b]{0.3}}
\put(50,-8){\makebox(1,1)[b]{1}}
\put(70,-8){\makebox(1,1)[b]{3}}
\put(90,-8){\makebox(1,1)[b]{10}}
\put(-9,20){\makebox(1,.5)[l]{20}}
\put(-9,40){\makebox(1,.5)[l]{40}}
\put(-9,60){\makebox(1,.5)[l]{60}}
\put(-9,80){\makebox(1,.5)[l]{80}}

\put(5,90){\large $F_{mp}$(\%)}
\put(40,-15){Beam energy (GeV)}

\end{picture}
\end{center}
\caption{
Efficiency ($\bullet$) of crystal-assisted slow extraction,
and beam lost in nuclear interactions in crystal ($\star$)
and on the aperture ($\otimes$),
as functions of beam kinetic energy.
}
  \label{slow}
\end{figure}
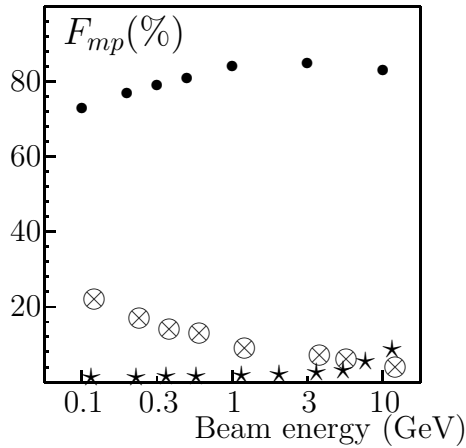

However, simulations (Figure 3)
show that, although the loss for nuclear interaction in thin crystal
at lower energy
becomes insignificant, the extraction efficiency is limited
by another factor -
multiple scattering to the aperture
(set in these simulations at 0.8 times the bending angle).
This factor is dominating with decreasing energy, Figure 3.
Nonetheless, the efficiency of crystal-assisted extraction
is still 75-80\% at a few hundred MeV.
In these simulations we assumed the bending angle of 2 mrad at any energy,
and described particle revolutions in the ring by linear matrices
with parameters chosen from the Indiana University CIS \cite{cis} -
just for illustrative purpose, despite the very broad energy range considered.
The crystal size was scaled linearly with energy as suggested
by the physics.
Besides the two kinds of beam loss shown in the picture,
nuclear interactions in the crystal ($\star$) and
multiple scattering to the aperture ($\otimes$),
there was a dechanneling loss, 7-8\% of the beam total
in the cases considered, due to the particles
channeled only part of the crystal and respectively bent just part
of the 2 mrad. The dechanneled particles are then lost somewhere
on the machine apertures, similarly to multiply scattered particles.


The divergence of the extracted beam in the horizontal plane (plane of
bending) is as small as $\pm$0.15 mrad$\times (pv)^{-1/2}$
(full width, where $pv$ is beam momentum times velocity in GeV)
and defined by crystal properties only.
In vertical plane, the scattering in the encounters with crystal contributes
(about 1 mrad at 250 MeV),
to be added quadratically to the initial divergence.
The experience in crystal extraction also shows
that this method may have
other benefits
such as flat time structure of the extracted beam.

\section{Conclusions}

Efficient systems for slow extraction and halo scraping
at 0.1-10 GeV accelerators and storage rings
can be designed on the base of bent crystals.
Beam particles can be steered at the angles of several mrad
with efficiencies of 70-90\%.

\section{Acknowledgements}

I am very much indebted to Jie Wei, Wu-Tsung Weng, Dejan Trbojevic,
and Walter Scandale
for useful discussions and support.

\end{document}